\documentclass[twocolumn, tighten]{aastex6}

\usepackage{color}
\usepackage{CJK}

\newcommand{\te}{t_{\rm E}}
\newcommand{\thetae}{\theta_{\rm E}}
\newcommand{\pie}{\pi_{\rm E}}

\definecolor{darkbrown}{RGB}{139,69,19}

\shorttitle{OGLE-2016-BLG-0613}
\shortauthors{Han et al.}
\begin{document}

\title{OGLE-2016-BLG-0613LAB\lowercase{b}: A Microlensing Planet in a Binary System}

\author{
C.~Han\altaffilmark{01}, A.~Udalski\altaffilmark{02,21}, Gould, A.\altaffilmark{03,04,05,22,23}, 
C.-U.~Lee\altaffilmark{03,06,22}, Y.~Shvartzvald\altaffilmark{07,23,25}, 
W.~C.~Zang\altaffilmark{08,09,24}, S.~Mao\altaffilmark{08,10,11,24}, 
S.~Koz{\l}owski\altaffilmark{02,21}, \\
and\\
M.~D.~Albrow\altaffilmark{12}, S.-J.~Chung\altaffilmark{03,06}, K.-H.~Hwang\altaffilmark{03}, 
Y.~K.~Jung\altaffilmark{13}, D.~Kim\altaffilmark{01}, H.-W.~Kim\altaffilmark{03}, 
Y.-H.~Ryu\altaffilmark{03}, I.-G.~Shin\altaffilmark{13}, J.~C.~Yee\altaffilmark{13}, 
W.~Zhu$^{04}$,       
S.-M.~Cha\altaffilmark{03,14},S.-L.~Kim\altaffilmark{03,06}, D.-J.~Kim\altaffilmark{03}, 
Y.~Lee\altaffilmark{03,14}, B.-G.~Park\altaffilmark{03,06},   \\ 
(The KMTNet Collaboration),\\
J.~Skowron\altaffilmark{02}, P.~Mr{\'o}z\altaffilmark{02},  P.~Pietrukowicz\altaffilmark{02}, 
R.~Poleski\altaffilmark{02,04}, M.~K.~Szyma{\'n}ski\altaffilmark{02},  
I.~Soszy{\'n}ski\altaffilmark{02},  K.~Ulaczyk\altaffilmark{02}, M.~Pawlak\altaffilmark{02}\\
(The OGLE Collaboration),\\
C.~Beichman\altaffilmark{15}, G.~Bryden\altaffilmark{07}, S.~Calchi~Novati\altaffilmark{16}, 
B.~S.~Gaudi\altaffilmark{04},
C.~B.~Henderson\altaffilmark{15}, S.~B.~Howell\altaffilmark{17}, S.~Jacklin\altaffilmark{18}, \\
(The UKIRT Microlensing Team),\\
M.~T.~Penny\altaffilmark{04,26,23}, P.~Fouqu\'e\altaffilmark{19,20}, T.~S.~Wang\altaffilmark{08}\\
(CFHT-K2C9 Microlensing Collaboration)\\
}

\altaffiltext{01} {Department of Physics, Chungbuk National University, Cheongju 28644, Republic of Korea}
\altaffiltext{02} {Warsaw University Observatory, Al. Ujazdowskie 4, 00-478 Warszawa, Poland}
\altaffiltext{03} {Korea Astronomy and Space Science Institute, Daejon 34055, Republic of Korea}
\altaffiltext{04} {Department of Astronomy, Ohio State University, 140 W. 18th Ave., Columbus, OH 43210, USA}
\altaffiltext{05} {Max Planck Institute for Astronomy, K\"onigstuhl 17, D-69117 Heidelberg, Germany}
\altaffiltext{06} {Korea University of Science and Technology, 217 Gajeong-ro, Yuseong-gu, Daejeon 34113, Republic of Korea}
\altaffiltext{07} {Jet Propulsion Laboratory, California Institute of Technology, 4800 Oak Grove Drive, Pasadena, CA 91109, USA}
\altaffiltext{08} {Physics Department and Tsinghua Centre for Astrophysics, Tsinghua University, Beijing 100084, China}
\altaffiltext{09} {Department of Physics, Zhejiang University, Hangzhou, 310058, China}
\altaffiltext{10} {National Astronomical Observatories, Chinese Academy of Sciences, A20 Datun Rd., Chaoyang District, Beijing 100012, China}
\altaffiltext{11} {Jodrell Bank Centre for Astrophysics, Alan Turing Building, University of Manchester, Manchester M13 9PL, UK}
\altaffiltext{12} {University of Canterbury, Department of Physics and Astronomy, Private Bag 4800, Christchurch 8020, New Zealand}
\altaffiltext{13} {Smithsonian Astrophysical Observatory, 60 Garden St., Cambridge, MA, 02138, USA}
\altaffiltext{14} {School of Space Research, Kyung Hee University, Yongin 17104, Republic of Korea}
\altaffiltext{15} {IPAC/NExScI, Mail Code 100-22, Caltech, 1200 E. California Blvd., Pasadena, CA 91125}
\altaffiltext{16} {IPAC, Mail Code 100-22, Caltech, 1200 E. California Blvd., Pasadena, CA 91125}
\altaffiltext{17} {Kepler \& K2 Missions, NASA Ames Research Center, PO Box 1,M/S 244-30, Moffett Field, CA 94035}
\altaffiltext{18} {Vanderbilt University, Department of Physics \& Astronomy, Nashville, TN 37235, USA}
\altaffiltext{19} {CFHT Corporation, 65-1238 Mamalahoa Hwy, Kamuela, Hawaii 96743, USA}
\altaffiltext{20} {Universit\'e de Toulouse, UPS-OMP, IRAP, Toulouse, France}
\footnotetext[21]{The OGLE Collaboration.}
\footnotetext[22]{The KMTNet Collaboration.}
\footnotetext[23]{The UKIRT Microlensing Team.}
\footnotetext[24]{The CFHT Microlensing Collaboration.}
\footnotetext[25]{NASA Postdoctoral Program Fellow.}
\footnotetext[26]{Sagan Fellow.}

\begin{abstract}
We present the analysis of OGLE-2016-BLG-0613, for which the lensing light curve appears 
to be that of a typical binary-lens event with two caustic spikes but with a discontinuous 
feature on the trough between the spikes.  We find that the discontinuous feature was 
produced by a planetary companion to the binary lens.  We find 4 degenerate 
triple-lens solution classes, each composed of a pair of solutions according to the well-known
wide/close planetary degeneracy. One of these solution classes is excluded due to its relatively
poor fit. For the remaining three pairs of solutions, the most-likely primary mass is about
$M_1\sim 0.7\,M_\odot$ while the planet is a super-Jupiter. In all cases the system lies in the
Galactic disk, about half-way toward the Galactic bulge. However, in one of these three solution
classes, the secondary of the binary system is a low-mass brown dwarf, with relative mass ratios
(1 : 0.03 : 0.003), while in the two others the masses of the binary components are comparable.
These two possibilities can be distinguished in about 2024 when the measured lens-source relative
proper motion will permit separate resolution of the lens and source.
\end{abstract}


\section{Introduction}
 
More than half of stars belong to binary or multiple systems \citep{Abt1983}. From 
high-resolution imaging observations of host stars of {\it Kepler} extrasolar planets, 
\citet{Horch2014} concluded that the overall binary fraction of the 
planet-host stars is similar to the rate for field stars, suggesting that planets in 
binary systems are likely to be as common as those around single stars. The environment 
of the protoplanetary disk around a star in a binary system would have been affected 
by the gravitational influence of the companion and thus planets in binary systems 
are expected to be formed through a different mechanism from that of single stars 
\citep{Thebault2015}. However, the current leading theories about the planet formation 
such as the core-accretion theory \citep{Ida2004} and the disk instability theory 
\citep{Boss2006} have been mostly focused on single stars, and thus many aspects about 
the formation mechanism of planets in binary systems remain uncertain.  In order to 
derive a clear understanding of the formation mechanism, a sample comprising a large 
number of planetary systems in various types of binary systems will be important.

Microlensing can provide a tool to detect planets in binaries.  The method is important 
because it can detect planets that present significant difficulties for other major 
planet detection methods. Since the microlensing phenomenon occurs regardless of the 
light from lensing objects, it enables one to detect planets around faints stars or 
even dark objects \citep{Mao1991,  Gould1992b}. 
While the transit method is sensitive to circumbinary planets, 
wherein the planet orbits both stars in a close binary, and the high-resolution imaging 
method is sensitive to circumstellar planets, wherein the planet orbits one star of a 
very wide binary system, the microlensing method can detect both populations of circumbinary 
\citep{Han2008} and circumstellar planets \citep{Lee2008, Han2016}.

However, identifying microlensing planets in binary systems is a difficult task due 
to the complexity of triple-lensing light curves and the resulting complication in 
the analysis.  When a lensing event is produced by a single mass, the resulting light 
curve has a simple smooth and symmetric shape which is described by 3 lensing parameters 
of the time of the closest approach of the source to the lens, $t_0$, which defines the 
time of the peak magnification, the lens-source separation at that time,  $u_0$ (impact 
parameter), which determines the peak magnification of the event, and the Einstein time 
scale $\te$, which characterizes the duration of the event.  When a lens is composed of 
multiple components, on the other hand, the lensing light curve becomes greatly complex.

\begin{figure}
\includegraphics[width=\columnwidth]{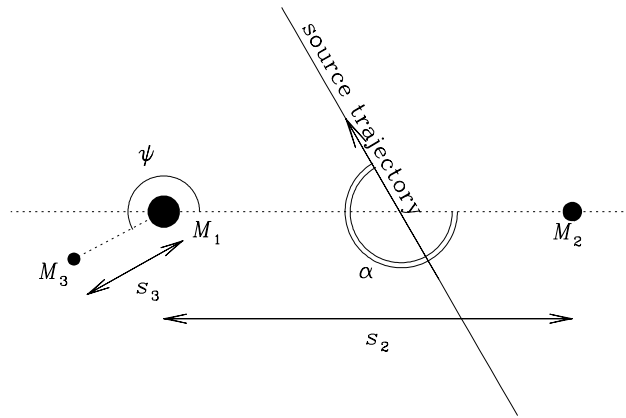}
\caption{
Graphical presentation of the triple-lensing parameters. 
}
\label{fig:one}
\end{figure}

The first cause of the light curve complexity is the increase of the parameters needed to 
describe the light curve. In order to describe the light curves produced by a binary lens, 
one needs 3 more parameters in addition to the single-lens parameters.  These include the 
separation $s_2$ and mass ratio $q_2=M_2/M_1$ between the binary lens components and the 
source trajectory angle $\alpha$ with respect to the binary axis.  Here $M_1$ and $M_2$ 
denote the masses of the primary and companion of the binary, respectively.  For a 
triple-lens system, such as the case of a planet in a binary system, one needs to add 3 
more parameters including the orientation angle $\psi$ of the third body $M_3$ and the 
separation $s_3$ and mass ratio $q_3$ between $M_1$ and $M_3$.  
In Figure~\ref{fig:one}, we provide the graphical presentation of the triple-lensing 
parameters used in our analysis. 
With the increased 
number of parameters, the parameter space to be explored in the analysis greatly 
increases, making the analysis of triple-lens system difficult.

The second major cause 
of the light curve complexity is the formation of caustics.  Caustics refer to curves on 
the source plane at which lensing magnifications of a point source become infinite, and 
thus the lensing light curves produced by source star's crossing over the caustic are 
characterized by sharp spikes.  Caustics of a binary-lens systems form closed curves, 
in which the topology of caustic curves depends on the binary separation and mass ratio 
\citep{Schneider1986, Bozza1999, Dominik1999a}.  The addition of a third component to the 
lens system greatly increases the complexity of the caustic topology, and the loops of 
caustics may overlap and intersect, resulting in nested curves \citep{Gaudi1998}. 
As a result, the topology of the triple lens system has not yet been fully understood 
although there have been some studies on the caustics of a subset of triple lens systems, 
e.g., \citet{Danek2015} and \citet{Luhn2016}.

Although very complex, microlensing light curves of some triple-lens systems can be 
readily analyzed. One such case is a planet in a binary system for which the masses 
of the stellar lens components overwhelm that of the planet, i.e.\ $M_1 \sim M_2 \gg M_3$. 
In this case, the overall shape of the lensing light curve is approximated by the 
binary-lens light curve of the $M_1$--$M_2$ pair, and the signal of the third body 
can be treated as a perturbation to the binary-lens curve.

There have been 5 triple lensing events published to date.  These include
OGLE-2006-BLG-109 \citep[two planet system,][]{Gaudi2008, Bennett2010},
OGLE-2012-BLG-0026 \citep[two planet system,][]{Han2013,Beaulieu2016},
OGLE-2013-BLG-0341 \citep[circumstellar planetary system,][]{Gould2014},
OGLE-2008-BLG-092 \citep[circumstellar planetary system,][]{Poleski2014}, and
OGLE-2007-BLG-349 \citep[circumbinary planetary system,][]{Bennett2016}.
For OGLE-2006-BLG-109, the caustics induced by the two planets interfere each 
other and the resulting caustic pattern is complex, making the analysis difficult. 
For the other 4 cases, however, the interference is minimal and thus the resulting 
caustic can be approximated by the superposition of the caustics induced by the 
individual companions \citep{Han2005}. This enables to treat $M_1$--$M_2$ and 
$M_1$--$M_3$ pairs as independent binary systems, making the analysis simplified.

In addition, there are two cases in the literature of lens systems that were 
originally identified as triples and then were later recognized to be binary 
lenses: MACHO-97-BLG-41 \citep{Bennett1999,Albrow2000,Jung2013,Ryu2017} and 
OGLE-2013-BLG-0723 \citep{Udalski2015a, Han2016}. In both cases, the ``additional'' 
caustic that had been thought to be caused by an additional body (whose properties 
were then evaluated based on the above-mentioned principle of superposition) was 
actually a minor-image caustic that had moved during the event due to the orbital motion 
of the binary.  
The first case is particularly instructive. 
Soon after the triple model was introduced, \citet{Albrow2000} had developed an 
alternative (binary) model from the analysis based on a completely different data 
set which had additional data points and thus provided stronger constraints
on the first caustic compared to the \citet{Bennett1999} study.
Later, \citet{Jung2013} combined these 
data sets and showed that the binary model was strongly favored. Subsequently, 
\citet{Ryu2017} showed that the fundamental reason for the superiority of the binary 
model is that the orbtial-motion of the binary was already strongly present in the 
brief encounter of the source with the central caustic.
In brief, triple lenses inhabit a wide range of parameter space, leading to 
considerably different levels of difficulty in analyzing them and disentangling 
them from binary lenses.  While considerable progress has been made, much is still 
being learned.

In this work, we report a planet in a binary system that was 
detected from the analysis of the microlensing event OGLE-2016-BLG-0613. The overall 
light curve of the event appears to be consistent with that of a typical caustic-crossing 
binary-lens event with two strong spikes, but it exhibits a short-term discontinuous 
feature on the smooth "U"-shape trough region between the caustic spikes. We find that 
the short-term feature is produced by a planet-mass companion to the binary lens.

The paper is organized as follows.  In Section 2, we discuss 
the observation of the lensing event and the data acquired from it.  In 
Section 3, we describe the procedures of modeling the observed 
lensing light curve and estimate the physical parameters of the lens system.  In 
Section4, we discuss the scientific importance of the 
lensing event.  We summarize the results and conclude in Section 5.

\section{Data}

The microlensing event OGLE-2016-BLG-0613 occurred on a star located toward the 
Galactic bulge field. The equatorial coordinates of the source star are 
$(\alpha,\delta)_{\rm J2000}=(17^{\rm h} 57^{\rm m} 02^{\rm s}\hskip-2pt .50,
-28^\circ 06' 58''\hskip-2pt .2)$.  The corresponding Galactic coordinates are 
$(l,b)=(1^\circ \hskip-2pt .99,-1^\circ \hskip-2pt .74)$, which is very close 
to the Galactic center. The event was discovered by the Early Warning System 
\citep{Udalski2003} of the Optical Gravitational Lensing Experiment 
\citep[OGLE:][]{Udalski2015b} survey, which monitors the Galactic bulge field using 
the 1.3m Warsaw telescope at the Las Campanas Observatory in Chile.  The discovery 
of the event was announced on 2016 April 11.  Most of the OGLE data were taken 
in the standard Cousins $I$ band and some $V$-band images were taken for color 
measurement.

\begin{figure}
\includegraphics[width=\columnwidth]{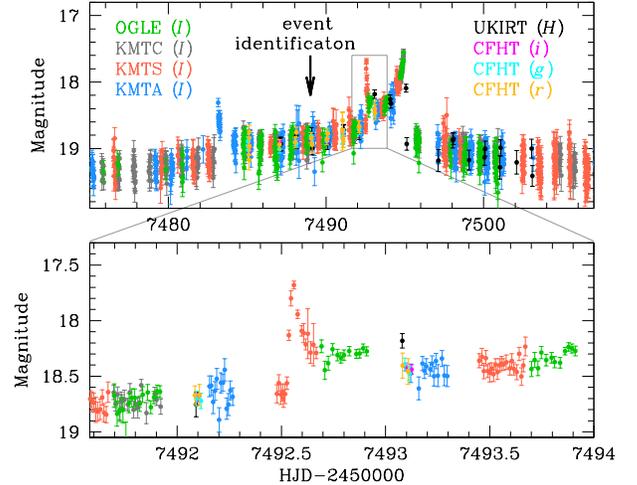}
\caption{
Light curve of OGLE-2016-BLG-0613. The lower panel shows the enlarged view of 
the short-term discontinuous anomaly that occurred at ${\rm HJD}\sim 2457493$. 
The arrow in the upper panel denotes the time when the event was identified.  
}
\label{fig:two}
\end{figure}

The event was also in the field of the Korea Microlensing Telescope Network 
\citep[KMTNet:][]{Kim2016} survey, which monitors the bulge field using 3 globally 
distributed telescopes located at the Cerro Tololo Interamerican Observatory in 
Chile (KMTC), the South African Astronomical Observatory in South Africa (KMTS), 
and the Siding Spring Observatory in Australia (KMTA). The aperture of each 
telescope is 1.6m.  The camera, which is composed of 4 chips, provides a 
4-deg$^{2}$ field of view.  The event OGLE-2016-BLG-0613 was in one of the 
three major fields toward which observations were conducted with 15-minute cadence.  
For these major fields, the KMTNet survey conducts alternating observations with 
6-arcminute offset in order to cover the gaps between chips of the camera.  As a 
result, the KMTNet data are composed of two sets (denoted by ``BLG02'' and ``BLG42'').  
KMTNet observations were also conducted in the standard Cousins $I$ band with 
occasional observations in $V$ band.

The event was also observed by the surveys conducted using the 3.8m United Kingdom 
Infrared Telescope \citep[UKIRT Microlensing survey:][]{Shvartzvald2017} and 
the 3.6m Canada France Hawaii Telescope (CFHT). Both telescopes are located at the 
Mauna Kea Observatory in Hawaii.  UKIRT observations were conducted in $H$ band and 
the images from CFHT observations were taken in $i$, $r$, and $g$ bands.

In Figure~\ref{fig:two}, we present the light curve of 
OGLE-2016-BLG-0613.  At the time of being identified 
(${\rm HJD}'={\rm HJD}-2450000\sim 7489$), the light curve of the event already 
showed deviations from the smooth and symmetric form of a point-mass event.  
With the progress of the event, the light curve exhibited a ``U''-shape trough, which 
is a characteristic feature that appears when a source star passes through the inner 
region of the caustic formed by a binary lens.  Caustics of a binary lens form closed 
curves, and thus caustic crossings occur in pairs.  The first caustic spike, which 
occurred at ${\rm HJD}'\sim 7483$, was inferred from the trough feature of the OGLE 
data and identified later by the KMTA data.  The second caustic spike occurred at 
${\rm HJD}'\sim 7495$.  The ``baseline object'' of the event was very faint with a baseline 
magnitude $I_{\rm OGLE}\sim 19.4$.  Furthermore, the baseline of the light curve 
showed a systematic trend of declination. As a result, the event drew little attention 
and thus it was not covered by follow-up observations.

\begin{figure}
\includegraphics[width=\columnwidth]{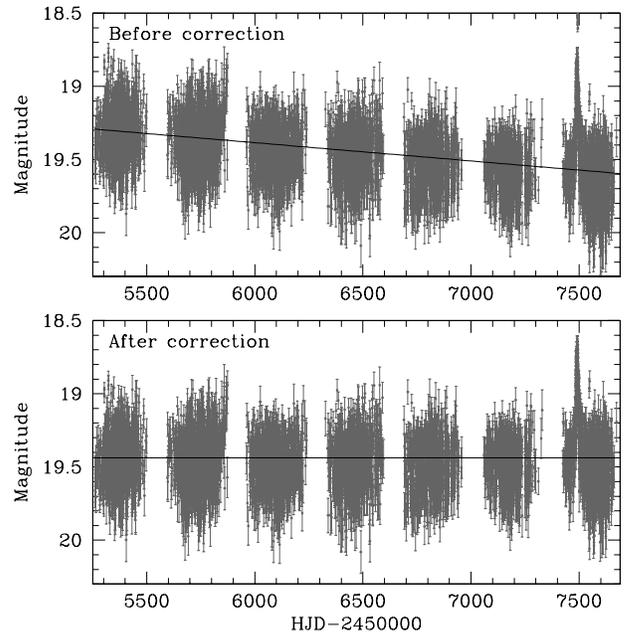}
\caption{
Systematic trend in the OGLE baseline data. The upper and lower panels show the data 
before and after the correction, respectively.
}
\label{fig:three}
\end{figure}

Preliminary analysis of the event was done as a part of real-time modeling efforts 
that were conducted to check the scientific importance of anomalous events. From 
this, it was noticed that there exists a short-term discontinuous feature at 
${\rm HJD}'\sim 7493$ in a U-shape trough region between the caustic spikes. 
We present the enlarged view of the anomaly in the lower panels of 
Figure~\ref{fig:two}.  Successive modeling of the light curve 
conducted with the progress of the event yielded solutions that can describe the 
overall light curve.  However, the short-term anomaly could not be explained by 
models based on the binary-lens interpretation.  See more discussion in 
Section 3.

Photometry of the images taken from observations were conducted by using various 
versions of codes developed based on the the difference imaging analysis method \citep[DIA:][]{Alard1998}.  
The OGLE data were processed with the customized pipeline \citep{Udalski2003}.  
The UKIRT data, which were taken in $H$ band, were also reduced with the DIA technique. 
The data taken by the KMTNet and CFHT surveys were reduced with customized versions of PySIS 
\citep{Albrow2009} and ISIS, respectively.

\begin{deluxetable}{ccrrc}
\tablecaption{Error bar correction factors \label{table:one}}
\tablewidth{0pt}
\tabletypesize{\footnotesize}
\tablehead{
\multicolumn{2}{c}{Data set} &
\multicolumn{1}{c}{$N$}  &
\multicolumn{1}{c}{$k$}  &
\multicolumn{1}{c}{$\sigma_{\rm min}$}  
}
\startdata                                              
OGLE      &                 & 1924  &  1.68  &   0.001     \\     
KMTC      & (BLG02 Field)   &  292  &  1.45  &   0.001     \\ 
--        & (BLG42 Field)   &  363  &  1.46  &   0.001     \\     
KMTS      & (BLG02 Field)   &  675  &  1.51  &   0.001     \\      
--        & (BLG42 Field)   &  630  &  1.93  &   0.001     \\       
KMTA      & (BLG02 Field)   &  397  &  1.40  &   0.001     \\    
--        & (BLG42 Field)   &  383  &  1.34  &   0.001     \\          
UKIRT     &                 &  71   &  1.41  &   0.020     \\    
CFHT      & ({\it i} band)  &  44   &  1.43  &   0.020     \\      
--        & ({\it r} band)  &  47   &  1.78  &   0.020     \\  
--        & ({\it g} band)  &  45   &  1.25  &   0.020      
\enddata                                              
\end{deluxetable}

As mentioned, the baseline of the event exhibits a systematic trend by which the 
baseline magnitude gradually increases. See the upper panel of 
Figure~\ref{fig:three}, which shows the 7 year baseline 
since 2010. Such a trend is often produced by a blended star that is moving away from 
the source star, e.g. OGLE-2013-BLG-0723 \citep{Han2016}. As the blend moves away, 
less flux is included within the tapered aperture of photometry for the source flux 
measurement, causing declining baseline. We remove the trend by conducting a linear 
fit to the baseline. See the light curve after the baseline correction presented in 
the lower panel of Figure~\ref{fig:three}.

For the analysis of the data taken from different telescopes and processed using 
different photometry codes, we readjust error bars of each data set following 
the usual procedure described in \citet{Yee2012}, i.e.
\begin{equation}
\sigma=k (\sigma_0^2 + \sigma_{\rm min}^2)^{1/2}.
\label{eq1}
\end{equation}
Here $\sigma_0$ denotes the error bar estimated from the photometry pipeline, 
$\sigma_{\rm min}$ is a factor used to make the error bars be consistent 
with the scatter of data, and $k$ is the scaling factor used to make $\chi^2/{\rm dof}=1$. 
The adopted values of the error-bar correction factors $k$ and $\sigma_{\rm min}$ are 
listed in Table~\ref{table:one}. Also presented is the number 
of data points, $N$, for the individual data sets.

\begin{figure}
\includegraphics[width=\columnwidth]{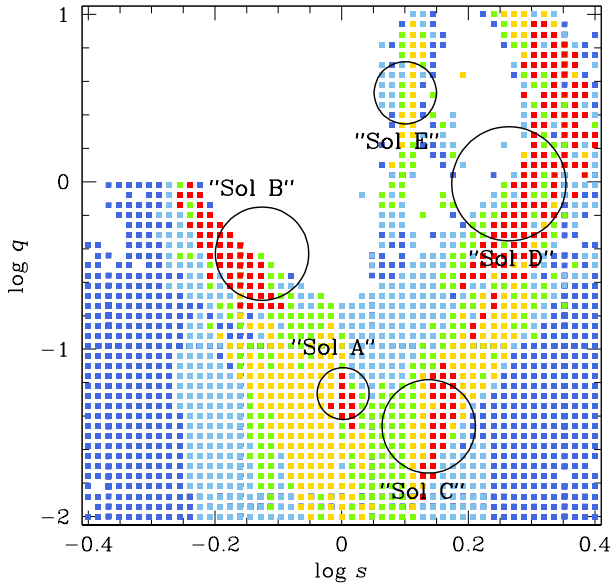}
\caption{
$\Delta\chi^2$ map in the $\log s$--$\log q$ parameter space.  
The encircled regions represent the locations of the 5 local binary-lens solutions.
The color coding represent points in the MCMC chain within 
$1n\sigma$ (red),
$2n\sigma$ (yellow),
$3n\sigma$ (green),
$4n\sigma$ (cyan), and
$5n\sigma$ (blue), where $n=8$.
}
\label{fig:four}
\end{figure}

\section{Analysis}

\subsection{Binary-Lens Modeling}

Since the sharp spikes in the lensing light curve are characteristic features 
of caustic-crossing binary-lens events, we start modeling of the light curve 
based on the assumption that the lens is composed of two masses, $M_1$ and $M_2$. 
For the simple case in which the relative lens-source motion is rectilinear, 
the light curve of a binary-lens event is described by 7 geometric parameters plus 
2 parameters representing the fluxes from the source, $F_s$, and blend, $F_b$, 
for each data set.  The geometric lensing parameters include 3 of a single-lens 
event ($t_0$, $u_0$ and $\te$), another 3 parameters describing the binary lens 
($s$, $q$, and $\alpha$), and the ratio of the angular source radius $\theta_*$ to 
the angular Einstein radius $\thetae$, i.e.\ $\rho=\theta_*/\thetae$ (normalized 
source radius).  The lengths of $u_0$ and $s$ are normalized to $\theta_{\rm E}$.

\begin{figure}
\epsscale{0.95} \plotone{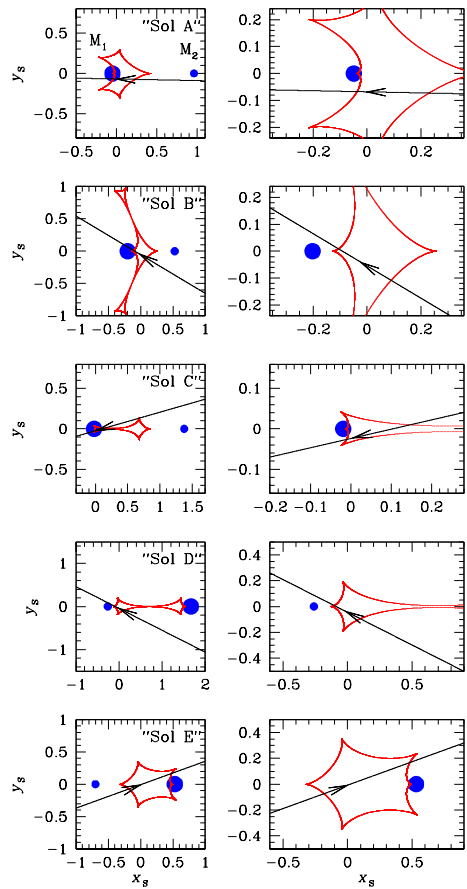}
\caption{
Caustic geometry of the 5 local binary-lens solutions.  For each local, the left 
panel shows the whole view and the right panel shows the zoom of the caustic-crossing 
region.  In each panel, the cuspy closed curve represents the caustic and the line 
with an arrow is the source trajectory.  The blue dots represent the lens components 
where the bigger and smaller dots denote the primary ($M_1$) and companion ($M_2$), 
respectively.
}
\label{fig:five}
\end{figure}

In the preliminary binary-lens modeling, we first conduct a grid search over 
the parameter space of $s$, $q$, and $\alpha$, while the remaining parameters 
($t_0$, $u_0$, $\te$, and $\rho$) are searched for using a downhill approach.  We 
choose $s$, $q$, and $\alpha$ as the grid parameters because lensing magnifications 
vary sensitively to the small changes of these parameters.  For the
downhill approach, we use the Markov Chain Monte Carlo (MCMC) method.  The initial 
values of the MCMC parameters are roughly guessed considering the characteristics 
of the light curve such as the duration, caustic crossing times, etc.  From the 
preliminary search, we find that (1) there exist multiple local minima that can 
describe the overall feature of the light curve but (2) none of these solutions 
can explain the short-term anomaly at ${\rm HJD}'\sim 7493$.

\begin{deluxetable*}{lccccc}
\tablecaption{Local binary-lens solutions \label{table:two}}
\tablewidth{0pt}
\tabletypesize{\footnotesize}
\tablehead{
\multicolumn{1}{c}{Parameters} &
\multicolumn{1}{c}{`Sol A'}    & 
\multicolumn{1}{c}{`Sol B'}    &
\multicolumn{1}{c}{`Sol C'}    & 
\multicolumn{1}{c}{`Sol D'}    & 
\multicolumn{1}{c}{`Sol E'}      
}
\startdata
$\chi^2$             &          9212.4                 &          9185.7          &        9179.2          &           9183.6        &         9195.1           \\  
$t_0$ (HJD')         &  7493.443 $\pm$ 0.055           &  7490.032 $\pm$ 0.053    & 7493.872 $\pm$ 0.054   &  7490.004 $\pm$ 0.081   &  7486.311 $\pm$ 0.109    \\
$u_0$                &     0.069 $\pm$ 0.001           &     0.046 $\pm$ 0.001    &    0.023 $\pm$ 0.001   &     0.039 $\pm$ 0.001   &    -0.006 $\pm$ 0.007    \\
$\te$ (days)         &    44.63r $\pm$ 0.39            &    52.87  $\pm$ 0.66     &   74.09  $\pm$ 0.20    &    53.39  $\pm$ 0.40    &    17.15  $\pm$ 0.20     \\
$s$                  &     1.011 $\pm$ 0.006           &     0.730 $\pm$ 0.006    &    1.393 $\pm$ 0.003   &     1.926 $\pm$ 0.008   &     1.228 $\pm$ 0.006    \\
$q$                  &     0.050 $\pm$ 0.002           &     0.382 $\pm$ 0.005    &    0.026 $\pm$ 0.001   &     1.002 $\pm$ 0.081   &     6.032 $\pm$ 0.002    \\
$\alpha$ (rad)       &     3.157 $\pm$ 0.010           &     3.681 $\pm$ 0.009    &    2.915 $\pm$ 0.008   &     3.611 $\pm$ 0.012   &    -0.348 $\pm$ 0.016    \\
$\rho$ ($10^{-3}$)   &     0.21  $\pm$ 0.07            &     0.21  $\pm$ 0.04     &    0.31  $\pm$ 0.04    &     0.31  $\pm$ 0.05    &     0.31  $\pm$ 0.02     
\enddata                                              
\tablecomments{${\rm HJD}'={\rm HJD}-2450000$.}
\end{deluxetable*}

It is a well known fact that there can exist multiple degenerate solutions in binary 
lensing modeling due to deep symmetries in the lens equation \citep{Griest1998,
Dominik1999b, An2005}. There is also at least one well-studied ``accidental'' degeneracy
between binary light curves due to planetary-mass and roughly equal-mass binaries, which 
does not occur for any deep reason \citep{Choi2012, Bozza2016}.  However, the general 
problem of ``accidental'' degeneracies is poorly studied. One general principle, however, 
is that the lower the quality and quantity of data, the wider is the range of possible 
binary geometries that may fit the data reasonably well.  As seen from Figure~\ref{fig:two}, 
OGLE-2016-BLG-0613 was exceptionally faint for a planet-bearing microlensing event, 
never getting brighter than $I\sim 17.5$. Moreover, although it was overall densely 
covered, by chance both the binary caustic entrance (${\rm HJD}^\prime\sim 7483$) and 
exit (${\rm HJD}^\prime\sim 7495$) fell in the gap between KMTC and KMTA coverage.  
Hence, there is a total of only one data point (from UKIRT) on these features. Binary 
events whose caustics lack such coverage are rarely, if ever, intensively studied, so 
that little is known about how they are impacted by ``accidental'' degeneracies. 
Therefore, one must pay particular attention to identifying all degenerate topologies.

Although several local solutions are found from the preliminary grid search, some local 
solutions might be still missed possibly due to a poor guess of the initial values of 
the MCMC parameters or some other reasons.  We, therefore, check the existence of additional 
local solutions using two systematic approaches.

In the first approach, we conduct a series of additional grid searches in which 
we provide various combination of MCMC parameters as initial values.  For a single 
lensing event, the values of the MCMC parameters are well characterized by the peak 
time (for $t_0$), peak magnification (for $u_0$), and the duration of the event 
(for $\te$).  For OGLE-2016-BLG-0613, however, it is difficult to estimate these 
values from the light curve and thus it might be that some local solutions 
have been missed if the given initial parameters were too far away from the correct 
ones.  From these searches, we identify 5 local solutions. Among them, 2 local solutions 
were missed in the initial search mainly due to the large difference 
between the Einstein time scales given as an initial value and the recovered value 
from modeling.

In the second approach, we directly consider each of 21 different caustic topologies that are
consistent with the overall morphology of the light curve. As seen in 
Figure~\ref{fig:two}, the caustic structures appear near the 
peak of the pre-caustic plus post-caustic light curve. The caustic structure must
therefore be either a four-sided (``central'') caustic or a six-sided (``resonant'') caustic. Allowing for
the symmetry of these caustic structures around the binary axis, there are $(4/2)\times (4-1)=6$
topologically distinct source paths for the central caustic and $(6/2)\times(6-1)=15$ for the resonant
caustic. We seed each of these topologies with an arbitrary $(s,q)$ geometry (that permits such a
path) and set initial values of $(t_0,u_0,t_{\rm E},\alpha)$ such that the caustic entrance and exit
occur at approximately the correct times. We then allow all parameters to vary using $\chi^2$
minimization. Although the initial seed solutions generally provide extremely poor matches to the
data, the derived local minima are always in rough accord with the data, showing that the
approach is working. Nevertheless, only four of these solutions have $\chi^2$ within a few
hundred of the global minimum. These yield the same five solutions found by the grid searches
above, with one of the four topology solutions corresponding to a close-wide pair of grid-search
solutions. See below.

In Figure~\ref{fig:four}, we present the locations of the local 
solutions in the $\Delta\chi^2$ map of the $\log s$ -- $\log q$ parameter space.  The 
individual local solutions are marked by circles and labeled as `Sol A', `Sol B', `Sol C', 
`Sol D', and `Sol E', respectively.  In Table~\ref{table:two}, we 
present the lesing parameters of the individual local binary-lensing solutions.  In 
Figure~\ref{fig:five}, we also present the geometry of 
the lens systems, which show the source trajectory with respect to the positions of the 
lens components and caustics.  One finds that the lensing parameters of the local solutions 
span over wide ranges.  For example, the range of the binary mass ratio is 
$0.03 \lesssim q \lesssim 6.0$.\footnote{We note that $q>1.0$ implies that the source 
approaches closer to the lower-mass component of the binary lens.}
 This implies that the observed light curve is serendipitously  
described by multiple local solutions with widely different lensing parameters.

Although the overall shape of the observed light curve is described by
multiple solutions, none of the solutions can explain the short-term
anomaly at ${\rm HJD}'\sim 7493$.
The anomaly is unlikely to be caused by systematics in the data because
the feature was covered commonly by the OGLE, KMTS, KMTA, and UKIRT
data. The region between caustic spikes can deviate from a smooth U shape
if the source within the caustic asymptotically approaches the caustic curve.
In such a case, however, the deviation occurring during the caustic approach
is smooth, while the observed short-term anomaly appears to be a small-scale
caustic-crossing feature composed of both caustic entrance
(at ${\rm HJD}'\sim 7292.5$) and exit (${\rm HJD}'\sim 7293.1$) and the
U-shape trough between them.

\begin{figure}
\includegraphics[width=\columnwidth]{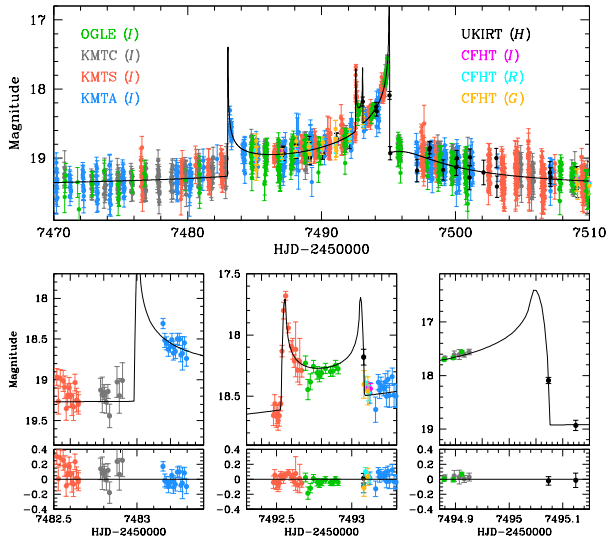}
\caption{
Model light curve of a triple-lens solution. 
The upper panel shows the model fit for the overall light curve, while
the lower 3 panels show the fits for the caustic-crossing regions.  
We note that the presented model light curve is that of `Sol B' and
that `Sol C (wide)' and `Sol D' provide models with similar fits. 
}
\label{fig:six}
\end{figure}

\begin{figure*}
\epsscale{0.75} \plotone{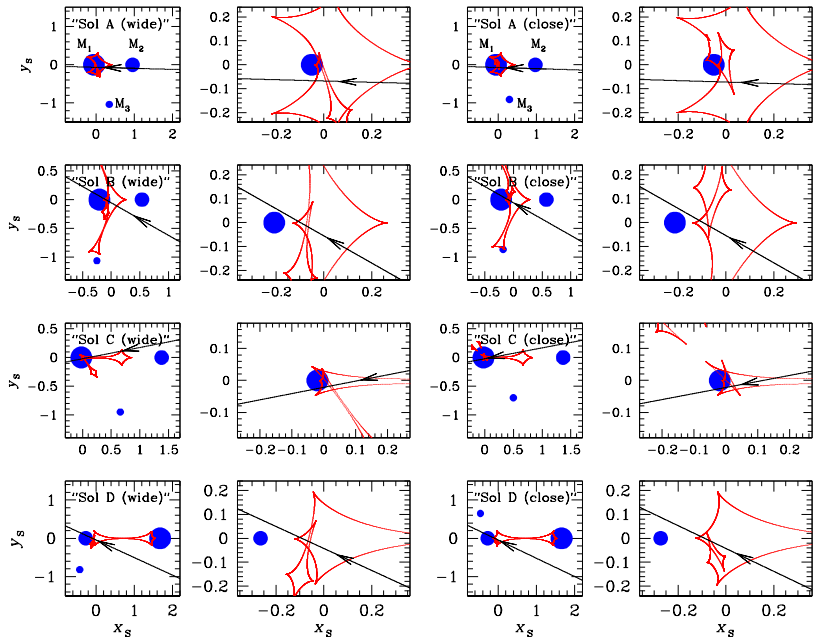}
\caption{
Caustic geometry of the 4 local triple-lens solutions. For each of the 
solutions, marked by `Sol A', `Sol B', `Sol C', and `Sol D', a pair of 
solutions, which are marked by `close' and  `wide', resulting from the 
close/wide degeneracy of the third body are presented. 
Notations are same as those in Fig.~\ref{fig:five}. 
}
\label{fig:seven}
\end{figure*}

\subsection{Triple-Lens Modeling}

Binary lenses form closed curves, it is mathematically impossible for a binary lens
to generate two successive caustic entrances without an 
intervening caustic exit, nor similarly two successive caustic exits without an 
intervening caustic entrance. In the present case, we have both two such successive 
entrances followed by two such successive exits. This suggests that one should consider 
a new interpretation of the event other than the binary-lens interpretation.

It is known that a third body of a lens can cause caustic curves to be self-intersected 
\citep{Schneider1992, Petters2001, Danek2015, Luhn2016}, and the light curve resulting 
from the source trajectory passing over the intersected part of the caustic can result 
in an additional caustic-crossing feature within the major caustic feature. We
therefore conduct a triple-lens modeling of the observed light curve in order to check 
whether the short-term anomaly can be explained by a third body.

The 3-body lens modeling is conducted in two steps.
In the first step, we conduct a grid search for the parameters related to the third 
body ($s_3$, $q_3$, and $\psi$) by fixing the binary-lens parameters at the values 
obtained from the binary-lens modeling.
Once approximate values of the third-body parameters are found, we then refine the 
3-body solution by allowing all parameters to vary.
The first step is based on the assumption that the overall light curve is well 
described by a binary-lens model and the signal of the third body can be treated as a 
perturbation to the binary-lens curve. For OGLE-2016-BLG-0613, this assumption is valid 
because of the good binary-lens fit to the overall light curve and the short-term nature 
of the anomaly. 
The lensing parameters related to the third  body include the separation $s_3$ and the 
mass ratio $q_3=M_3/M_1$ between the third and the primary of the binary lens and the 
position angle of the third body measured from the binary axis, $\psi$.  According to 
our definition of the position angle $\psi$, the third body is located at 
$(x_{\rm L,3},y_{\rm L,3})=(x_{\rm L,1}+s_3 \cos \psi, y_{\rm L,1}+s_3 \sin \psi)$, 
where $(x_{\rm L,1},y_{\rm L,1})$ represents the position of the primary of the binary 
lens, i.e.\ $M_1$.

From triple-lens modeling, we find that the short-term anomaly can be explained by 
introducing a low-mass third body to the binary-lens solutions `Sol A' through `Sol D'.
For the case of `Sol E', we find no triple-lens solution that can explain the short-term 
anomaly.  For each of the solutions that can explain the short-term anomaly, we find a 
pair of triple-lens solutions resulting from the close/wide degeneracy of the third-body, 
i.e.\ $s_3$ versus (approximately) $s_3^{-1}$ \citep{Han2013, Song2014}.  
We designate the solutions with $s_3>1$ 
and $s_3<1$ as ``wide'' and ``close'' solutions, respectively.\footnote{Note that
for ``Sol D'', the ``wide'' solution has $s_3<1$. In fact,
it requires some care to map the simple symmetries found
by \citet{Griest1998} for a single host, to the present case
of a binary host. In particular, one sees from Figures~4 and 6
that the planet is basically perturbing the magnification field
of the primary, whereas the lensing parameters are defined relative
to the mass of the entire system. If $s_3$ were rescaled to the
mass of the primary [as in, e.g., \citet{Gould2014}], then the
two values would be $s_3=1.16$ (wide) and $s_3=0.93$ (close).}

In Table~\ref{table:three}, we list the lensing parameters of 
the triple-lens solutions along with $\chi^2$ values.  Also presented are the source 
and blend magnitudes, $I_s$ and $I_b$, respectively.\footnote{We note that the error 
in the blend, $I_b$, is not presented because 
the error is formally extremely small, 
less than 0.001 mag. This estimate is accurate in the sense that $F_b = F_{\rm base}-F_s$, 
where $F_{\rm base}$ is the flux due to the nearest ``star'' in the DoPhot-based catalog 
derived from the template image.  However, this quantity is itself the result of complex 
processing of a crowded-field image and does not precisely correspond to any physical 
quantity. Within the context of the analysis, it is just a nuisance parameter, although 
it can in principle place upper limits on light from the lens.  }
From the mass ratios of the third body, one finds 
that $1.8\times 10^{-3} \lesssim q_3 \lesssim 6.4\times 10^{-3}$, indicating that the 
third body is a planetary-mass object regardless of the models.  From the mass ratios 
between the binary components, one finds that $q_2 \lesssim 0.06$ for `Sol A' and 
`Sol C', while $q_2 \gtrsim 0.35$ for `Sol B' and `Sol D'.  This indicates that the 
lens of `Sol A' and `Sol C' would be composed of a star, a brown dwarf (BD), and a planet, 
while the lens of `Sol B' and `Sol D' would consist of a stellar binary plus a planet.

The model light curve of the triple-lens solution and the residual from the observed 
data are presented in Figure~\ref{fig:six} for the `Sol B (wide)' 
model as a representative model.  We note that $\chi^2$ differences among the 
models `Sol B', `Sol C' and `Sol D' are $\lesssim  30$ and thus the fits of `Sol C' 
and `Sol D' are similar to the fit of the presented model.  For `Sol A', on the other 
hand, the fit is worse than the presented model by $\Delta\chi^2 \sim 80$.  
Figure~\ref{fig:seven} shows the caustic geometry of the 
individual triple-lens solutions.  In all cases, it is found that the short-term anomaly 
is produced by the deformation of the caustic caused by the third body.

We check whether the model further improves by additionally considering the parallax effect 
induced by the orbital motion of the Earth around the sun \citep{Gould1992a}. We find that 
the microlens parallax  $\pi_{\rm E}$ can be neither reliably measured nor 
meaningfully constrained. 
We note that the event was in the field of the space-based lensing survey using 
the {\it Kepler} space telescope \citep[$K2$C9:][]{Henderson2016}.
Since the {\it Kepler} telescope is in a heliocentric orbit, the space-based 
observations could have led to the measurement of the microlens parallax
\citep{Refsdal1966, Gould1994}.
The $K2$C9 campaign was planned to start at ${\rm HJD}'\sim 7486$, when the 
event was in progress. However, the campaign could start only at ${\rm HJD}'\sim 7501$
because of an emergency mode and thus the event was missed.

\subsection{Source Star}

We characterize the source star by measuring the de-reddened color and brightness, that 
are calibrated  using the centroid of giant clump (GC) in the color-magnitude diagram (CMD) 
\citep{Yoo2004}.  Although the event was observed in $V$ band by both the OGLE and KMTNet 
surveys, the $V$-band photometry quality is not good enough for a reliable measurement of 
the $V$-band baseline source flux due to the faintness of the source star combined with 
the high extinction toward the field.  We therefore use the OGLE $I$-band data and UKIRT 
$H$-band data.

\begin{deluxetable*}{lrrrrrrrr}
\tablecaption{Local triple-lens solutions \label{table:three}}
\tablewidth{0pt}
\tabletypesize{\footnotesize}
\tablehead{
\multicolumn{1}{c}{Parameters} &
\multicolumn{2}{c}{`Sol A'}    &
\multicolumn{2}{c}{`Sol B'}    &
\multicolumn{2}{c}{`Sol C'}    & 
\multicolumn{2}{c}{`Sol D'}    \\
\multicolumn{1}{c}{}           &
\multicolumn{1}{c}{Wide}           &
\multicolumn{1}{c}{Close}           &
\multicolumn{1}{c}{Wide}           &
\multicolumn{1}{c}{Close}           &
\multicolumn{1}{c}{Wide}           &
\multicolumn{1}{c}{Close}           &
\multicolumn{1}{c}{Wide}           &
\multicolumn{1}{c}{Close}           
}
\startdata
$\chi^2$                 &       4869.9        &        4881.05        &        4799.2        &        4812.9         &          4789.2         &        4805.6        &       4802.4         &        4816.8         \\
$t_0$ (HJD')             &       7493.551      &        7493.572       &        7490.135      &        7490.489       &          7494.153       &        7494.177      &       7490.095       &        7489.920       \\
                         & $\pm$ 0.038         &  $\pm$ 0.043          & $\pm$  0.063         & $\pm$  0.120          &   $\pm$  0.048          & $\pm$  0.017         & $\pm$ 0.075          &  $\pm$ 0.039          \\
$u_0$                    &       0.068         &        0.072          &        0.048         &        0.047          &          0.021          &        0.022         &       0.038          &        0.038          \\
                         & $\pm$ 0.001         &  $\pm$ 0.001          & $\pm$  0.001         & $\pm$  0.001          &   $\pm$  0.001          & $\pm$  0.001         & $\pm$ 0.002          &  $\pm$ 0.001          \\
$\te$ (days)             &       44.48         &        42.13          &       51.45          &       44.94           &         74.62           &       71.90          &      53.53           &       52.01           \\
                         & $\pm$ 0.71          &  $\pm$ 0.19           & $\pm$  0.41          & $\pm$  1.33           &   $\pm$  1.69           & $\pm$  0.57          & $\pm$ 0.43           &  $\pm$ 0.10           \\
$s_2$                    &       1.009         &        1.025          &        0.743         &        0.802          &          1.396          &        1.386         &       1.941          &        1.932          \\
                         & $\pm$ 0.004         &  $\pm$ 0.004          & $\pm$  0.005         & $\pm$  0.015          &   $\pm$  0.010          & $\pm$  0.004         & $\pm$ 0.019          &  $\pm$ 0.003          \\
$q_2$                    &       0.053         &        0.055          &        0.386         &        0.359          &          0.029          &        0.027         &       1.051          &        1.114          \\
                         & $\pm$ 0.001         &  $\pm$ 0.001          & $\pm$  0.010         & $\pm$  0.011          &   $\pm$  0.002          & $\pm$  0.001         & $\pm$ 0.052          &  $\pm$ 0.019          \\
$\alpha$ (rad)           &       3.169         &        3.172          &        3.659         &        3.663          &          2.948          &        2.954         &       3.572          &        3.587          \\
                         & $\pm$ 0.009         &  $\pm$ 0.012          & $\pm$  0.005         & $\pm$  0.004          &   $\pm$  0.010          & $\pm$  0.003         & $\pm$ 0.014          &  $\pm$ 0.005          \\
$s_3$                    &       1.111         &        0.971          &        1.064         &        0.872          &          1.168          &        0.872         &       0.833          &        0.677          \\
                         & $\pm$ 0.001         &  $\pm$ 0.001          & $\pm$  0.001         & $\pm$  0.002          &   $\pm$  0.008          & $\pm$  0.005         & $\pm$ 0.012          &  $\pm$ 0.004          \\
$q_3$ ($10^{-3}$)        &       2.44          &        2.07           &        5.54          &        4.43           &          3.27           &        3.24          &       6.39           &        1.77           \\
                         & $\pm$ 0.07          &  $\pm$ 0.01           & $\pm$  0.13          & $\pm$  0.13           &   $\pm$  0.24           & $\pm$  0.18          & $\pm$ 0.49           &  $\pm$ 0.05           \\
$\psi$ (rad)             &       5.079         &        5.079          &        4.672         &        4.774          &          5.332          &        5.350         &       4.519          &        1.852          \\
                         & $\pm$ 0.010         &  $\pm$ 0.007          & $\pm$  0.011         & $\pm$  0.022          &   $\pm$  0.016          & $\pm$  0.004         & $\pm$ 0.011          &  $\pm$ 0.011          \\
$\rho$ ($10^{-3}$)       &       0.35          &        0.39           &        0.30          &        0.36           &          0.22           &        0.23          &       0.32           &        0.25           \\
                         & $\pm$ 0.03          &  $\pm$ 0.03           & $\pm$  0.03          & $\pm$  0.03           &   $\pm$  0.02           & $\pm$  0.02          & $\pm$ 0.03           &  $\pm$ 0.02           \\              
$I_{s,{\rm OGLE}}$ (mag) & 21.84               &  21.81                & 22.25                & 22.01                 &   23.00                 & 22.90                &  21.96               &  21.96                \\
                         & $\pm$ 0.01          &  $\pm$ 0.01           & $\pm$ 0.02           & $\pm$ 0.04            &   $\pm$ 0.05            & $\pm$ 0.02           &  $\pm$ 0.02          &  $\pm$ 0.01           \\
$H_{s}$ (mag)            & 18.55               &  18.52                & 18.96                & 18.72                 &   19.71                 & 19.61                &  18.67               &  18.67                \\
                         & $\pm$ 0.04          &  $\pm$ 0.04           & $\pm$ 0.04           & $\pm$ 0.06            &   $\pm$ 0.06            & $\pm$ 0.04           &  $\pm$ 0.04          &  $\pm$ 0.04           \\
$I_{b,{\rm OGLE}}$ (mag) & 19.58               &  19.59                &  19.55               & 19.56                 &   19.50                 & 19.51                &  19.57               &   19.57            
\enddata                                               
\tablecomments{${\rm HJD}'={\rm HJD}-2450000$.}
\end{deluxetable*}

\begin{figure}
\includegraphics[width=\columnwidth]{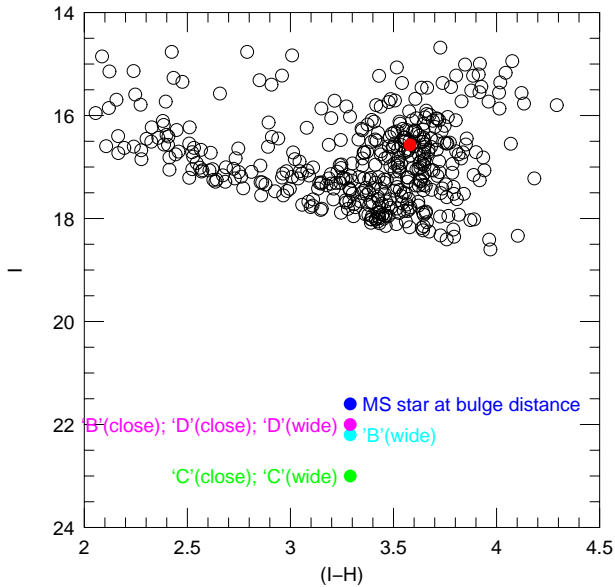}
\caption{
Source locations with respect to the giant clump centroid (red dot) in the 
$(I-H)/H$ color-magnitude diagram of stars around the source.  We mark 4 source 
locations, where the blue dot represents the location of a late G-type main-sequence 
star at a bulge distance, the magenta dot denotes the source location with the 
brightness estimated from `Sol B (close)', `Sol D (close)', and `Sol D (wide)', 
and the cyan dot is the source location with the source brightness estimated from 
`Sol B (wide)', and finally the green dot is the source location with the brightness 
estimated from `Sol C (close)' and `Sol C (wide)'.
}
\label{fig:eight}
\end{figure}

In Figure~\ref{fig:eight}, we mark the location of the source star 
with respect to the GC centroid in the $(I-H)/I$ CMD of neighboring stars around the source. 
We measure the GC centroid
$(I-H,I)_{\rm GC} = (3.58,16.57)\pm (0.01,0.07)$, where
the errors are derived from the standard error of the mean of
clump stars.  As is virtually always the case, for each model,
the source color is essentially identical $(I-H)_s = 3.29\pm 0.04$
to much greater precision than the measurement error, which is
dominated by the $H$-band photometric errors.  Thus
$\Delta(I-H) = -0.29\pm 0.04$.  Transformation to $(V-I)$ of
both the value and error using the color-color relation
of \citet{Bessell1988} yields $\Delta(V-I) = -0.21\pm 0.03$.
Based on the well-defined de-reddened color of GC 
centroid $(V-I)_{0,{\rm GC}}=1.06$ \citep{Bensby2011}, it is estimated that the de-reddened 
color of the source is
\begin{equation}
(V-I)_{0,s}=(V-I)_{0,{\rm GC}} + \Delta(V-I) 
= 0.85\pm 0.03.
\label{eq2}
\end{equation} 
This color corresponds to that of a late G-type main-sequence star with an 
absolute magnitude of $M_{I,s}\sim 4.9$. From the absolute magnitude of the GC centroid 
of $M_{I,{\rm GC}}=-0.12$ \citep{Nataf2013} combined with its apparent brightness $I_{\rm GC}=16.57$ 
measured on the CMD, then, the apparent magnitude of the source star should be 
\begin{equation}
I_s=I_{\rm GC}-(M_{I,{\rm GC}}-M_{I,s}) \sim 21.6.
\label{eq3}
\end{equation}

\subsection{Partial Resolution of Triple-Lens Degeneracy}

It is found that the observed light curve can be explained by $(4\times 2=)8$ degenerate
triple-lens solutions. In this subsection, we further investigate the individual
solutions in order to check the feasibility of resolving the degeneracy among
the solutions.

First, we exclude `Sol A' due to its relatively poor fit to the observed light 
curve compared to the other solutions.  From the comparison of $\chi^2$ values  
presented in Table~\ref{table:three}, it is found that 
`Sol A' provides a fit that is poorer than `Sol B', `Sol C', and `Sol D' by 
$\Delta\chi^2=70.7$, 80.7, and 67.5, respectively. These $\chi^2$ differences 
are statistically significant enough to exclude `Sol A'.

This leaves three pairs of solutions, one pair for each of `Sols B, C, D'. `Sol C' 
(wide) is favored by $\Delta\chi^2=10$ over any other solutions. However, the source 
position on the CMD is a priori substantially less likely than for `Sol B' and `Sol D'.  
The source brightness determined from the source flux $F_s$, presented in 
Table~\ref{table:three},
are $I_s=23.00\pm 0.05$ and 
$22.90\pm 0.02$ for the wide and close solutions, respectively.  
These are $\gtrsim 1.3$ magnitude 
fainter than $I_s=21.6$, i.e.\ Eq.~(\ref{eq3}), estimated based on the relative source 
position with respect to the GC centroid in the CMD.  See Figure~\ref{fig:eight}.  
To explain such a faint source would require either that the source is very distant 
(at e.g., $D_{\rm S}\sim 15\,$kpc), or that it is intrinsically dim for its color due to low 
metallicity (e.g., [Fe/H]$\sim -1.3$). Either of these is possible, although with low 
probability.  For example, \citet{Bensby2017a} find a total of 3 stars with [Fe/H]$<-1.2$ 
out of their sample of 90 microlensed bulge dwarfs and subgiants. 
Similarly, \citet{Ness2013} found about 2\% of stars
with ${\rm [Fe/H]}<-1.2$ in a much larger sample, albeit
substantially farther from the Galactic plane than typical
microlensing events.  Compare Figures 1 and 10 of \citet{Bensby2017b}
with Figures 1 and 6 of \citet{Ness2013}, respectively.
The \citet{Bensby2017a} 
sample is appropriate for comparison because it faces qualitatively similar selection 
biases to microlensing surveys that lead to planet detection. On the other hand, 
\citet{Bensby2017a} found no clear evidence for microlensed sources that lay substantially
 behind the Galactic bulge.

In brief `Sol C' is mildly favored by $\chi^2$ but requires a somewhat unlikely source. If its
$\Delta\chi^2=10$ advantage could be interpreted at face value according to Gaussian statistics,
this solution would be mildly preferred by $\Delta\ln L = \ln(3/90)+10/2 = 1.6$. However, it is
well known that microlensing light curves are affected by subtle systematics at the
$\Delta\chi^2=$few level, and so cannot be judged according to naive Gaussian statistics.
Therefore, we consider that `Sol C' is viable but somewhat disfavored.

On the other hand, `Sol B' and `Sol D' not only provide good fits to the observed 
light curve but also meet the source brightness constraint.  The apparent source 
magnitudes estimated from $F_s$ values of the models are 
$I_s = 22.25 \pm 0.02$/$22.01 \pm 0.04$
for the wide/close solutions of `Sol B' 
and 
$I_s= 21.96\pm 0.02$/$21.96 \pm 0.01$
for the wide/close solutions of `Sol D'.  These are in accordance with 
$I_s \sim 21.6$ 
estimated 
from the CMD.  We find that the lensing parameters of `Sol B' and `Sol D' models are in 
the relation of
\begin{equation}
s_B \sim {1+q_B \over 1-q_B \sqrt{1+q_D}} s_D^{-1},
\label{eq4}
\end{equation}  
where $(s_B,q_B)$ represent the binary separation and mass ratio of `Sol B' and $(s_D,q_D)$ 
represent those of `Sol D'.  This indicates that `Sol B' and `Sol D'  are the pair of 
solutions resulting from the well-known close/wide binary-lens degeneracy which is rooted 
in the symmetry of the lens equation and thus can be severe \citep{Dominik1999b, An2005}.  
For OGLE-2016-BLG-0613, the degeneracy is quite severe with $\Delta\chi^2\lesssim 5$.  
Note in particular from Figure~\ref{fig:five} that the topologies of `Sol B' and 
`Sol D' are essentially identical and are also distinct from the other three topologies shown.

\subsection{Angular Eintein Radius}

The angular Einstein radius is estimated from the combination of the normalized 
source radius $\rho$ and the angular source radius $\theta_*$, i.e.\ 
$\thetae=\theta_*/\rho$. The normalized source radius is measured by analyzing 
the caustic-crossing part of the light curve where the lensing magnifications 
are affected by finite-source effects.  Although three of the four caustic crossings 
were covered by either zero or one point (and so yield essentially no information 
about $\rho$), the planetary-caustic entrance was very well covered by KMTS data.  
See the lower middle panel of Figure~\ref{fig:six}.

The angular source radius is estimated based on the de-reddened color and brightness.
The de-reddeded $V$-band brightness is estimated by 
\begin{equation}
V_{0,s}=I_{0,s}+(V-I)_{0,s}, 
\label{eq5}
\end{equation}
where $I_{0,s} = I_{{\rm GC},0} - (I_{\rm GC} - I_s)$, $I_{{\rm GC},0} = 14.38$
\citep{Nataf2013}, 
$I_{\rm GC} = 16.57\pm 0.07$,
$(V-I)_{0,s} = 0.85\pm 0.03$ 
[Eq.~(\ref{eq2})], and $I_s$ is given for
each solution in Table~\ref{table:three}.
We convert $(V-I)_0$ into $(V-K)_0$ using the $VI/VK$ relation of \citet{Bessell1988}.
Then, the angular source radius $\theta_*$ is determined from the relation between 
$\theta_*$ and  $(V, V-K)_0$ provided by \citet{Kervella2004}.

\begin{deluxetable*}{lrrrrrr}
\tablecaption{Einstein ring radius and proper motion\label{table:four}}
\tablewidth{0pt}
\tabletypesize{\footnotesize}
\tablehead{
\multicolumn{1}{c}{Parameters} &
\multicolumn{2}{c}{`Sol B'}    &
\multicolumn{2}{c}{`Sol C'}    &
\multicolumn{2}{c}{`Sol D'}    \\
\multicolumn{1}{c}{}           &
\multicolumn{1}{c}{Wide}       &
\multicolumn{1}{c}{Close}     &
\multicolumn{1}{c}{Wide}       &
\multicolumn{1}{c}{Close}     &
\multicolumn{1}{c}{Wide}       &
\multicolumn{1}{c}{Close}           
}
\startdata
$\theta_{\rm E}$ (mas)  & $1.20 \pm 0.24$ & $1.12 \pm 0.22$  &   $1.15 \pm 0.22$  &   $1.15 \pm 0.22$     &  $1.28 \pm 0.25$  &   $1.63 \pm 0.32$   \\
$\mu$ (mas yr$^{-1}$)   & $8.50 \pm 1.71$ & $9.06 \pm 1.75$  &   $5.64 \pm 1.11$  &   $5.86 \pm 1.14$     &  $8.70 \pm 1.72$  &   $11.5 \pm 2.20$    
\enddata                                               
\end{deluxetable*}

\begin{deluxetable*}{lrrrrrr}
\tablecaption{Physical lens parameters \label{table:five}}
\tablewidth{0pt}
\tabletypesize{\footnotesize}
\tablehead{
\multicolumn{1}{c}{Parameters} &
\multicolumn{2}{c}{`Sol B'}    &
\multicolumn{2}{c}{`Sol C'}    &
\multicolumn{2}{c}{`Sol D'}    \\
\multicolumn{1}{c}{}           &
\multicolumn{1}{c}{Wide}       &
\multicolumn{1}{c}{Close}      &
\multicolumn{1}{c}{Wide}       &
\multicolumn{1}{c}{Close}      &
\multicolumn{1}{c}{Wide}       &
\multicolumn{1}{c}{Close}           
}
\startdata
$D_{\rm L}$ (kpc)       &   $3.41^{+1.34}_{-1.42}$   &  $3.56^{+1.30}_{-1.51}$  &   $3.19^{+1.40}_{-1.27}$      &  $3.21^{+1.35}_{-1.20}$       &   $3.84^{+1.44}_{-1.63}$        &    $2.93^{+1.59}_{-1.44}$     \\
$M_1$ ($M_\odot$)       &   $0.72^{+0.55}_{-0.42}$   &  $0.65^{+0.50}_{-0.37}$  &   $0.80^{+0.65}_{-0.43}$      &  $0.81^{+0.64}_{-0.44}$       &   $0.66^{+0.58}_{-0.38}$        &    $0.63^{+0.72}_{-0.41}$     \\    
$M_2$ ($M_\odot$)       &   $0.28^{+0.21}_{-0.16}$   &  $0.23^{+0.18}_{-0.13}$  &   $0.023^{+0.019}_{-0.012}$   &  $0.023^{+0.017}_{-0.012}$    &   $0.69^{+0.61}_{-0.40}$        &    $0.70^{+0.80}_{-0.46}$     \\    
$M_3$ ($M_J$)           &   $4.18^{+3.19}_{-2.43}$   &  $3.01^{+2.32}_{-1.71}$  &   $2.74^{+2.23}_{-1.51}$      &  $2.75^{+2.17}_{-1.49}$       &   $4.42^{+3.88}_{-2.54}$        &    $1.17^{+1.33}_{-0.76}$     \\    
$a_{\perp,1-2}$ (au)    &   $2.96^{+1.16}_{-1.23}$   &  $3.02^{+1.10}_{-1.28}$  &   $5.00^{+2.19}_{-1.99}$      &  $5.00^{+2.10}_{-1.87}$       &   $9.09^{+3.41}_{-3.86}$        &    $8.66^{+0.42}_{-0.38}$     \\    
$a_{\perp,1-3}$ (au)    &   $6.40^{+2.51}_{-2.63}$   &  $3.28^{+1.20}_{-1.39}$  &   $4.19^{+1.84}_{-1.67}$      &  $3.15^{+1.32}_{-1.18}$       &   $3.90^{+1.46}_{-1.66}$        &    $3.04^{+0.15}_{-0.13}$    
\enddata                                               
\end{deluxetable*}

In Table~\ref{table:four}, we present the angular Einstein 
radii for the viable models `Sol B', `Sol C', and `Sol D'.  Also presented are the 
relative lens-source proper motion determined by 
\begin{equation}
\mu = {\thetae \over t_{\rm E}},
\qquad
\thetae\equiv \sqrt{\kappa M \pi_{\rm rel}},
\label{eq6}
\end{equation}
where $\pi_{\rm rel} = {\rm au}(D_L^{-1}-D_S^{-1})$ is the lens-source relative parallax 
and $\kappa \equiv 4G/(c^2{\rm au})\simeq 8.14\ {\rm mas}\ M_\odot^{-1}$.  The inferred 
angular Einstein radii for the ensemble of solutions are in the range of 
$1.1 \lesssim \thetae/{\rm mas} \lesssim 1.6$.  These large values of $\thetae$ virtually 
rule out bulge lenses and so directly imply that the lens is in the Galactic disk. That 
is, if we hypothesized that the lens were in the bulge, then (since the bulge is a 
relatively old population), we could infer $M\la 1.3\,M_\odot$. Hence, for 
$\thetae\ga 1.1\,{\rm mas}$, Equation~(\ref{eq6}) implies $\pi_{\rm rel}\ga 0.11\,{\rm mas}$, 
which would contradict the hypothesis that the lens was in the bulge.

\begin{figure*}
\epsscale{0.85} \plotone{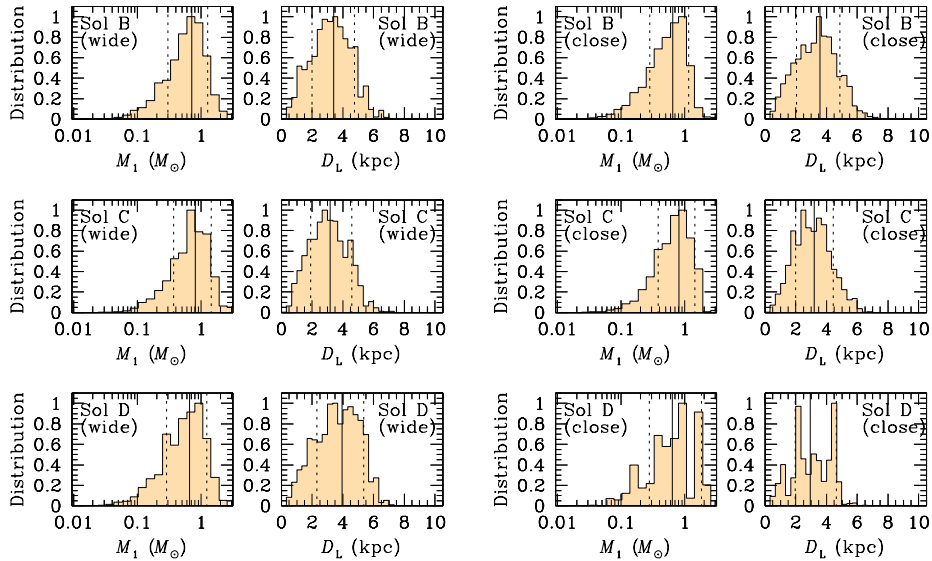}
\caption{
Distributions of the primary mass $M_1$ and the distance to the lens $D_{\rm L}$
for the individual solutions obtained from Bayesian analysis.
In each panel, the solid vertical line represents the median value 
and the two dotted lines represent the $1\sigma$ range of the distribution.
}
\label{fig:nine}
\end{figure*}

\subsection{Physical Lens Parameters}

For the unique determinations of the lens mass $M$ and distance $D_{\rm L}$, 
one needs to measure both the angular Einstein radius $\thetae$ and the 
microlens parallax $\pie$:
\begin{equation}
M = {\thetae\over\kappa\pie};
\qquad
D_{\rm L} = {{\rm au}\over\pie\thetae +\pi_{\rm S}},
\label{eq7}
\end{equation}
where $\pi_{\rm S}= {\rm au}/D_S$ is the source parallax.
For OGLE-2016-BLOG-0613, $\thetae$ 
is measured but $\pie$ is not measured and thus the values of $M$ and $D_{\rm L}$ 
cannot be uniquely determined.  However, one can still constrain the physical lens 
parameters based on the measured values of the event time scale $t_{\rm E}$ and the 
angular Einstein radius $\thetae$.

In order to estimate the mass and distance to the lens, we conduct a Bayesian 
analysis of the event based on the mass function combined with the models of 
the physical and dynamical distributions of objects in the Galaxy.  We use 
the initial mass function of \citet{Chabrier2003a} for the mass function of 
Galactic bulge objects, while  we use the present day mass function of 
\citet{Chabrier2003b} for disk objects.  In the mass function, we do not include 
stellar remnants, i.e.\ white dwarfs, neutron stars, and black holes, because it 
would be difficult for planets to survive the AGB/planetary-nebular phase of 
stellar evolution and no planet belonging to a remnant host is so far known, 
e.g.\ \citet{Kilic2009}.  For the matter distribution, we adopt the Galactic model 
of \citet{Han2003}. In this model, the matter density distribution is constructed 
based on a double-exponential disk and a triaxial bulge.  We use the dynamical model 
of \citet{Han1995} to construct the velocity distribution.  In this model, the disk 
velocity distribution is assumed to be Gaussian about the rotation velocity of the 
disk and the bulge velocity distribution is modeled to be a triaxial Gaussian with 
velocity components deduced from the flattening of the bulge via the tensor virial 
theorem.  Based on these models, we generate a large number of artificial lensing 
events by conducting a Monte Carlo simulation. We then estimate the ranges of $M$ 
and $D_{\rm L}$ corresponding to the measured event time scale and the angular 
Einstein radius.

In Table~\ref{table:five}, 
we list the physical parameters of the lens system estimated from the 
the Bayesian analysis.
In Figure~\ref{fig:nine}, we also present the distributions of the primary mass $M_1$
and the distance to the lens $D_{\rm L}$ 
obtained from the Bayesian analysis
for the individual solutions. 
The values denoted by $M_i$ represent the masses of the individual lens components, 
and $a_{\perp,1-2}$ and $a_{\perp,1-3}$ denote the projected $M_1-M_2$ and $M_1-M_3$ 
separations, respectively.  We note that the unit of $M_1$ and $M_2$ is the solar 
mass, $M_\odot$, while the unit of $M_3$ is the Jupiter mass, $M_J$.  The presented 
physical parameters are the median values of the corresponding distributions and the 
uncertainties are estimated as the standard deviations of the distributions.

All three pairs of solutions shown in Table~\ref{table:five}  
are comprised of a super-Jupiter planet in a
binary-star system, whose primary is a K dwarf. They are all located in the Galactic disk, about
half-way toward the bulge. The major difference among these solutions is that for `Sol B' and
`Sol D', the components of the binary are of equal (B) or comparable (D) mass, whereas for 
`Sol C', the second component of the binary is a low-mass brown dwarf.
We note that in all cases, the projected separations of the secondary and the planet are roughly
comparable, so that if the system lay in the plane of the sky it would be unstable. This is a
result of a generic selection bias of microlensing. Binary lenses, and in particular planets, 
are most easily discovered if they lie separated in projection by roughly one Einstein radius. 
See e.g., Figure~7 of \citet{Mroz2017}. This bias affects multi-lens systems by the square.
OGLE-2012-BLG-0026 \citep{Han2013} provides an excellent example of such bias. Hence, stable,
hierarchical systems are preferentially seen at an angle such that the projected separations are
comparable.

\section{Discussion}

OGLE-2016-BLG-0613 is of scientific importance because it demonstrates that planets in binary systems 
can be readily detected using the microlensing method. The planet is the fourth microlensing 
planet in binary systems followed by OGLE-2008-BLG-092L \citep{Poleski2014}, OGLE-2013-BLG-0341L 
\citep{Gould2014}, and OGLE-2007-BLG-349 \citep{Bennett2016}. Since the region of planet 
sensitivity is different from those of other planet-detection methods, the microlensing 
method will enrich the sample of planets in binaries, helping us to understand details 
about the formation mechanism of these planets.

The event illustrates the difficulty of 3-body lensing modeling. As shown in the previous 
section, interpreting the light curve suffers from multifold degeneracy due to the complexity 
of triple-lens topology combined with insufficient data quality. Such a difficulty in 
the interpretation was found in the case of another triple-lens event OGLE-2007-BLG-349 for 
which there existed two possible interpretations of the circumbinary-planet model and two-planet 
model. \citet{Bennett2016} were able to resolve the degeneracy with additional high-resolution 
images obtained from {\it Hubble Space Telescope} observations.

`Sol C' (star + brown-dwarf + super-Jupiter) represents a substantially different type of system
from either `Sol B' or `Sol D' (comparable mass binary with super-Jupiter). It would therefore be of
considerable interest to distinguish between these two classes. This will be quite straightforward
once the source and lens are sufficiently separated to be resolved in high-resolution images
(whether from space or the ground) because `Sol C' has both a much lower proper motion 
(Table~\ref{table:four}) 
and a much fainter source star (Figure~\ref{fig:eight}). 
In fact, even if `Sol C' is the correct solution, it is
only necessary to wait until the lens and source {\it would be} separately resolved in `Sol B' and
`Sol D', based on their substantially larger proper motions. In that case, if the lens and source
are not separately resolved, this non-detection would demonstrate that `Sol C' was correct. We
note that for the case of OGLE-2005-BLG-169, which had a similar proper motions to `Sol B' and
`Sol D', \citet{Batista2015} clearly resolved the source and lens with Keck adaptive optics
observations taken 8.2 years later, while \citet{Bennett2015} marginally resolved them after 6.5
years using the {\it Hubble Space Telescope}. In principle, it is also possible to both detect the
lens and measure its proper motion by subtracting out the source light from high-resolution
images. This may be possible in the present case. However, application of this approach would
be significantly complicated by the existence of several different solutions with very different
source fluxes.

\section{Conclusion}

We analyzed the  microlensing event OGLE-2016-BLG-0613 for which the light curve appeared 
to be that of a typical binary-lens event with two caustic spikes but with a short-term 
discontinuous feature on the smooth trough region between the spikes.  It was found that 
the overall feature of the light curve was described by multiple binary-lens solutions but 
the short-term discontinuous feature could be explained by none of these solutions.  We 
found that the discontinuous feature could be explained by introducing a low-mass planetary 
companion to the binary lens.  We found 4 degenerate triple-lens solutions, among which one 
was excluded due to the relatively poor fit compared to the other solutions. For each of the 
remaining three classes of solutions, there is a pair of sub-solutions according to the 
well-known close-wide degeneracy for planets. In two of the three classes of solutions, the 
two binary components are of comparable mass, while in the third, the second component of 
the binary is a low mass brown dwarf.  The degeneracy between the binary-star/planet lens 
model(s) and the star/brown-dwarf/planet lens model can be resolved by future high-resolution 
imaging observation.

\begin{acknowledgments}
Work by CH and DK was supported by the grant (2017R1A4A1015178) of
National Research Foundation of Korea. 
The OGLE project has received funding from the National Science Centre, Poland, 
grant MAESTRO 2014/14/A/ST9/00121 to A.~Udalski.  OGLE Team thanks Profs.\ M.~Kubiak, 
G.~Pietrzy{\'n}ski, and {\L}.~Wyrzykowski for their contribution to the collection of 
the OGLE photometric data over the past years.
A.~Gould and W.~Zhu acknowledges the support from NSF grant AST-1516842. 
We acknowledge the high-speed internet service (KREONET)
provided by Korea Institute of Science and Technology Information (KISTI).
Work by YS was supported by an appointment to the NASA Postdoctoral Program at the Jet Propulsion Laboratory,
California Institute of Technology, administered by Universities Space Research Association
through a contract with NASA.
The United Kingdom Infrared Telescope (UKIRT) is supported by NASA and
operated under an agreement among the University of Hawaii, the University
of Arizona, and Lockheed Martin Advanced Technology Center; operations are
enabled through the cooperation of the Joint Astronomy Centre of the Science
and Technology Facilities Council of the U.K.
We acknowledge the support from NASA HQ for the UKIRT observations in connection with $K2$C9.
Work by SM was supported by the National Science Foundation of China (11333003 and 11390372).
This research has made use of the KMTNet system operated by the Korea
Astronomy and Space Science Institute (KASI) and the data were obtained at
three host sites of CTIO in Chile, SAAO in South Africa, and SSO in
Australia.
\end{acknowledgments}

\end{document}